\documentclass[a4paper,10pt]{article}

\usepackage{epsf}
\usepackage{graphicx}    

\newcommand{\be}[1]{\begin{equation}\label{#1}}
\newcommand{\ee}{\end{equation}}
\newcommand{\bea}{\begin{eqnarray}}
\newcommand{\eea}{\end{eqnarray}}

\def\gsim{ \lower .75ex \hbox{$\sim$} \llap{\raise .27ex \hbox{$>$}} }
\def\lsim{ \lower .75ex \hbox{$\sim$} \llap{\raise .27ex \hbox{$<$}} }

\renewcommand{\markright}{\markright{\thepage}}

\begin{document}

\begin{titlepage}


\vspace{5mm}

\begin{center}

{\Large \bf Holographic dark energy in a cyclic universe}

\vspace{10mm}

{\large Jingfei Zhang,$^{1}$ Xin Zhang,$^{2}$ and Hongya Liu$^{1}$}

\vspace{5mm} {\em $^{1}$School of Physics and Optoelectronic
Technology, Dalian University of Technology, Dalian 116024, People's
Republic of China\\
$^{2}$Kavli Institute for Theoretical Physics China, Institute of
Theoretical Physics, Chinese Academy of Sciences (KITPC/ITP-CAS),
P.O.Box 2735, Beijing 100080, People's Republic of China}

\end{center}

\vspace{5mm}
\begin{abstract}
In this paper we study the cosmological evolution of the holographic
dark energy in a cyclic universe, generalizing the model of
holographic dark energy proposed by Li. The holographic dark energy
with $c<1$ can realize a quintom behavior; namely, it evolves from a
quintessence-like component to a phantom-like one. The holographic
phantom energy density grows rapidly and dominates the late-time
expanding phase, helping realize a cyclic universe scenario in which
the high energy regime is modified by the effects of quantum
gravity, causing a turnaround (and a bounce) of the universe. The
dynamical evolution of holographic dark energy in the regimes of low
energy and high energy is governed by two differential equations,
respectively. It is of importance to link the two regimes for this
scenario. We propose a link condition giving rise to a complete
picture of holographic evolution of a cyclic universe.

\end{abstract}

\end{titlepage}

\newpage

\setcounter{page}{2}

The astronomical observations over the past decade imply that our
universe is currently dominated by dark energy, which leads to an
accelerated expansion of the universe (see e.g. Refs.
\cite{SN,CMB,LSS}). The combined analysis of cosmological
observations suggests that the universe is spatially flat and
consists of about $70\%$ dark energy, $30\%$ dust matter (cold dark
matter plus baryons), and negligible radiation. The basic
characteristic of dark energy is that its equation of state
parameter $w$ (the definition of $w$ is $w=p/\rho$, where $\rho$ is
the energy density and $p$ is the pressure) has a negative value
($w<-1/3$). The most obvious candidate for dark energy is the
cosmological constant $\lambda$ \cite{Einstein:1917} for which
$w=-1$ (for reviews see e.g. Refs. \cite{cc}). However, the
cosmological constant always suffers from the ``fine-tuning'' and
``cosmic coincidence'' problems. Another candidate for dark energy
is the energy density associated with a dynamical scalar field; a
slowly varying, spatially homogeneous component. Typical examples of
such a type of dark energy are the so-called ``quintessence''
\cite{quintessence} and ``phantom'' \cite{phantom} models.
Quintessence dark energy provided by a canonical scalar field has an
equation of state $w>-1$, while the phantom energy associated with a
scalar field with a negative-kinetic energy has an equation of state
$w<-1$. It is remarkable that for phantom dark energy in this
scenario all the energy conditions in general relativity (including
the weak energy condition) are violated. Due to such a supernegative
equation of state, the phantom component leads to a ``big rip''
singularity at a finite future time, at which all bound objects will
be torn apart.

In the phantom scenario, generically, there exist two space-time
singularities in the universe; one is the initial ``big bang''
singularity, the other is the future ``big rip'' singularity. The
space-time singularities are disgusting for the majority of
theorists; thus a mechanism for avoiding the initial and future
singularities are attractive for physicists and cosmologists. An
effective way for eliminating the singularities is to introduce a
$\rho^2$ term with a negative sign to the Friedmann equation if the
energy is very high. Such a modified Friedmann equation with a
phantom energy component leads to a cyclic universe scenario in
which the universe oscillates through a series of expansions and
contractions. Phantom energy can dominate the universe today and
drive the current cosmic acceleration. Then, as the universe
expands, it becomes more and more dominant and its energy density
becomes very high. When the phantom energy density reaches a
critical value, a very high energy density, the universe reaches a
state of maximum expansion, which we call the ``turnaround point'',
and then it begins to recollapse, according to the modified
Friedmann equation. The contraction of the universe makes the
phantom energy density dilute away and the matter density dominate.
Once the universe reaches its smallest extent, the matter density
hits the value of the critical density, the modified Friedmann
equation leads to a ``bounce'', making the universe once again begin
to expand.

The idea of an oscillating universe was first proposed by Tolman in
the 1930's \cite{Tolman}. In recent years, Steinhardt, Turok and
collaborators \cite{cycsteinhardt} proposed a cyclic model of the
universe as an alternative to the inflation scenario, in which the
cyclicity of the universe is realized in the light of two separated
branes. The cyclic scenario discussed in this paper is distinguished
from the Steinhardt-Turok cyclic scenarios in that the phantom
energy plays a crucial role. In the oscillating or cyclic models,
the principal obstacles against success come from the problems of
black holes and entropy. For the discussions of the problems of
black holes and entropy in the cyclic scenario, see e.g. Refs.
\cite{phantombounce,bhphantom}.

Usually, the phantom energy density becomes infinite in a finite
time, leading to the big rip singularity. However, we expect that an
epoch of quantum gravity sets in before the energy density reaches
infinity. Therefore, we arrive at the notion that quantum gravity
governs the behavior of the universe both at the beginning and at
the end of the expanding universe, where the energy density is
enormously high. This high energy density physics may lead to
modifications to the Friedmann equation, such as in loop quantum
cosmology \cite{loop} and braneworld scenarios
\cite{Randall:1999,Shtanov:2002mb}, which causes the universe to
bounce when it is small, and to turn around when it is large. At
high energy densities, we employ the modified Friedmann equation
\begin{equation}
H^2={8\pi G\over 3}\rho\left(1-{\rho\over\rho_{\rm
c}}\right),\label{modiFeq}
\end{equation}
where $H=\dot{a}/a$ is the Hubble parameter, $G$ is the Newton
gravity constant, and $\rho_{\rm c}$ is the critical energy density
set by quantum gravity, distinguished from the usual critical
density $3M_{\rm pl}^2H^2$ (where $M_{\rm pl}=1/\sqrt{8\pi G}$ is
the reduced Planck mass). This modified Friedmann equation can be
derived from the effective theory of loop quantum cosmology
\cite{loop}, and also from the braneworld scenario
\cite{Shtanov:2002mb}. In loop quantum cosmology, the critical
energy density can be evaluated as $\rho_{\rm c}\approx
0.82\rho_{\rm pl}$, where $\rho_{\rm pl}=G^{-2}=2.22\times
10^{76}~{\rm GeV}^4$ is the Planck density. In the braneworld
scenario, $\rho_{\rm c}=2\sigma$, where $\sigma$ is the brane
tension, and a negative sign in Eq. (\ref{modiFeq}) can arise from a
second time-like dimension; but that gives difficulties with closed
time-like paths. In models motivated by the Randall-Sundrum scenario
\cite{Randall:1999}, the most natural energy scale of the brane
tension is of the order of the Planck mass, but the problem can
generally be treated for any value of $\sigma>{\rm TeV}^4$. Once the
energy density of the universe reaches the critical density
$\rho_{\rm c}$, the universe changes its evolution direction. At
that energy scale, if it has been expanding, it turns around and
begins to contract; if it has been contracting, it bounces and
begins to expand. Modifications to the Friedmann equation thus
motivate the bounce and the turnaround, both of which are
nonsingular.

Recently, considerable interest has been stimulated in explaining
the observed dark energy by the holographic dark energy model. The
holographic dark energy model is an attempt to probe the nature of
dark energy within the framework of a fundamental theory originating
from some considerations of the features of quantum gravity theory.
Concretely speaking, this model is constructed in the light of the
holographic principle \cite{holoprin} of quantum gravity. In the
holographic scenario, the dark energy is a dynamically evolving
vacuum energy density that can realize the phantom behavior. If the
holographic dark energy becomes a phantom, this scenario will
involve a big rip singularity in the far future unless the Friedmann
equation gets a quantum gravity correction in the high energy regime
as shown in Eq. (\ref{modiFeq}). Thus, a phantom-like holographic
dark energy can play a crucial role in realizing a cyclic universe
scenario, and inversely, a cyclic universe can endow the holographic
dark energy with peculiar features. In this paper, we shall study
the cyclic universe with a holographic phantom and investigate the
characteristic of the holographic dark energy in such a cyclic
universe.

According to the holographic principle, the number of degrees of
freedom for a system within a finite region should be finite and
should be bounded roughly by the area of its boundary. In the
cosmological context, the holographic principle will set an upper
bound on the entropy of the universe. Motivated by the Bekenstein
entropy bound, it seems plausible that one may require that for an
effective quantum field theory in a box of size $L$ with UV cutoff
$\Lambda$, the total entropy should satisfy $S=L^3\Lambda^3\leq
S_{BH}\equiv\pi M_{\rm pl}^2L^2$, where $S_{BH}$ is the entropy of a
black hole with the same size $L$. However, Cohen et al.
\cite{Cohen:1998zx} pointed out that to saturate this inequality
some states with Schwartzschild radius much larger than the box size
have to be counted in. As a result, a more restrictive bound, the
energy bound, has been proposed to constrain the degrees of freedom
of the system, requiring the total energy of a system with size $L$
not to exceed the mass of a black hole with the same size, namely,
$L^3\Lambda^4=L^3\rho_\Lambda\leq L M_{\rm pl}^2$. This means that
the maximum entropy is in the order of $S_{BH}^{3/4}$. When we take
the whole universe into account, the vacuum energy related to this
holographic principle is viewed as dark energy, usually dubbed
holographic dark energy. The largest IR cut-off $L$ is chosen by
saturating the inequality, so that we get the holographic dark
energy density
\begin{equation}
\rho_{\Lambda}=3c^2M_{\rm pl}^2L^{-2}~,\label{de}
\end{equation}
where $c$ is a numerical constant (note that $c>0$ is assumed), and
as usual $M_{\rm pl}$ is the reduced Planck mass. Hereafter, we will
use the unit $M_{\rm pl}=1$ for convenience. It has been conjectured
by Li \cite{Li:2004rb} that the IR cutoff $L$ should be given by the
future event horizon of the universe,
\begin{equation}
R_{\rm eh}(a)=a\int\limits_t^\infty{dt'\over
a(t')}=a\int\limits_a^\infty{da'\over Ha'^2}~.\label{eh}
\end{equation}
Such a holographic dark energy looks reasonable, since it may
provide simultaneously natural solutions to both dark energy
problems, as demonstrated in Ref. \cite{Li:2004rb}. The holographic
dark energy model has been tested and constrained by various
astronomical observations \cite{obs,obs2}. For other extensive
studies on the holographic dark energy, see e.g. Refs.
\cite{holoext1,holoext2}.

The holographic dark energy scenario reveals the dynamical nature of
the vacuum energy. When taking the holographic principle into
account, the vacuum energy density will evolve dynamically. The
dimensionless parameter $c$ plays a crucial role in the holographic
evolution of the universe. As has been pointed out in Refs.
\cite{obs}, the value of $c$ determines the destiny of the
holographic universe. When $c\geq 1$, the equation of state of dark
energy will evolve in the region of $-1\leq w\leq -1/3$. In
particular, if $c=1$ is chosen, the behavior of the holographic dark
energy will be more and more like a cosmological constant with the
expansion of the universe, such that ultimately the universe will
enter the de Sitter phase in the far future. When $c<1$, the
holographic dark energy will exhibit a quintom-like evolution
behavior (for ``quintom'' dark energy, see, e.g., Refs.
\cite{quintom} and references therein), i.e., the holographic
evolution will make the equation of state cross $w=-1$ (from $w>-1$
evolves to $w<-1$). That is to say, $c<1$ makes the holographic dark
energy today behave as a phantom energy that leads to a cosmic
doomsday (``big rip'') in the future. Nevertheless, as discussed
above, at high energy densities the Friedmann equation may be
modified to Eq. (\ref{modiFeq}) due to some possible quantum gravity
effects, which can successfully eliminate the big rip singularity.
It is remarkable that the analyses of the observational data imply
that the value of $c$ in the model of holographic dark energy is
very likely less than 1 \cite{obs}, i.e., the holographic dark
energy is very possibly behaving as a phantom energy presently.
Intriguingly, then, considering the modified Friedmann equation
(\ref{modiFeq}) in the high energy regime, the holographic dark
energy (with $c<1$) along with some dust-like matter components can
realize a cyclic universe scenario in which the cosmological
evolution is nonsingular.\footnote{In general, cyclic universe
models confront two severe problems making the infinite cyclicity
impossible. First, the black holes in the universe, which cannot
disappear due to the Hawking area theorems, grow ever larger during
subsequent cycles, and they eventually will occupy the entire
horizon volume during the contracting phase, so that the
calculations break down. The second problem is that the entropy of
the universe increases from cycle to cycle due to the second law of
thermodynamics, so that extrapolation into the past will lead back
to an initial singularity. In this paper, we do not consider these
difficulties.}

First, consider the low energy regime of the universe, $\rho\ll
\rho_{\rm c}$. At this regime, for the universe we have the usual
Friedmann-Robertson-Walker (FRW) case, $3H^2=\rho$. Consider a
universe filled with a matter component $\rho_{\rm m}$ (including
both baryon matter and cold dark matter) and a holographic dark
energy component $\rho_{\Lambda}$; then the Friedmann equation reads
\begin{equation}
3H^2=\rho_{\rm m}+\rho_{\Lambda}.\label{Feq}
\end{equation}
Defining the fractional densities $\Omega_\Lambda=\rho_\Lambda/3H^2$
and $\Omega_{\rm m}=\rho_{\rm m}/3H^2=\Omega_{\rm m}^0H_0^2H^{-2}
a^{-3}$, where $a$ is the scale factor of the universe and $a_0=1$
has been set, the Friedmann equation (\ref{Feq}) can also be
rewritten as
\begin{equation}
{H^2\over H_0^2}=\Omega_{\rm m}^0a^{-3}+\Omega_{\Lambda}{H^2\over
H_0^2}.\label{Feq2}
\end{equation}
Combining the definition of the holographic dark energy (\ref{de})
and the definition of the future event horizon (\ref{eh}), we derive
\begin{equation}
\int_a^\infty{d\ln a'\over Ha'}={c\over
Ha\sqrt{\Omega_{\Lambda}}}.\label{rh}
\end{equation} We notice that the Friedmann
equation (\ref{Feq2}) implies
\begin{equation}
{1\over Ha}=\sqrt{a(1-\Omega_{\Lambda})}{1\over H_0\sqrt{\Omega_{\rm
m}^0}}.\label{fri}
\end{equation} Substituting (\ref{fri}) into (\ref{rh}), one
obtains the following equation:
\begin{equation}
\int_x^\infty e^{x'/2}\sqrt{1-\Omega_{\Lambda}}dx'=c
e^{x/2}\sqrt{{1\over\Omega_{\Lambda}}-1},
\end{equation} where $x=\ln a$. Then taking the derivative with respect to $x$ in both
sides of the above relation, we easily get the dynamics satisfied by
the dark energy, namely the differential equation for the fractional
density of the dark energy,
\begin{equation}
\Omega'_{\Lambda}=\Omega_{\Lambda}(1-\Omega_{\Lambda})\left(1+{2\over
c}\sqrt{\Omega_{\Lambda}}\right),\label{deq}
\end{equation}
where the prime denotes the derivative with respect to $x=\ln a$.
This equation describes the behavior of the holographic dark energy
in the low energy regime completely, and it can be solved exactly
\cite{Li:2004rb}:
\begin{equation}
\ln\Omega_\Lambda-{c\over 2+c}\ln(1-\sqrt{\Omega_\Lambda})+{c\over
2-c}\ln (1+\sqrt{\Omega_\Lambda})-{8\over
4-c^2}\ln(c+2\sqrt{\Omega_\Lambda})=\ln a+y_0,\label{soldeq}
\end{equation}
where $y_0$ is an integration constant that can be determined by
setting today as an initial condition,
\begin{equation}
y_0=\ln(1-\Omega_{\rm m}^0)-{c\over 2+c}\ln(1-\sqrt{1-\Omega_{\rm
m}^0})+{c\over 2-c}\ln(1+\sqrt{1-\Omega_{\rm m}^0})-{8\over
4-c^2}\ln(c+2\sqrt{1-\Omega_{\rm m}^0}).\label{y0}
\end{equation}
From the energy conservation equation of the dark energy, the
equation of state of the dark energy can be given \cite{Li:2004rb}:
\begin{equation}
w=-1-{1\over 3}{d\ln\rho_{\Lambda}\over d\ln a}=-{1\over
3}(1+{2\over c}\sqrt{\Omega_{\Lambda}}).\label{w}
\end{equation} Note that the formula
$\rho_{\Lambda}=[\Omega_{\Lambda}/(1-\Omega_{\Lambda})]\rho_{\rm
m}^0a^{-3}$ and the differential equation of $\Omega_{\Lambda}$, see
(\ref{deq}), are used in the second equality sign.

As time passes by, the dark energy gradually dominates the evolution
of the universe, $\Omega_\Lambda$ increases to 1, and the most
important term on the left-hand side of (\ref{soldeq}) is the second
term; thus for large $a$, we get
\begin{equation}
\sqrt{\Omega_\Lambda}=1-2^{2+c\over 2-c}(2+c)^{8\over
c(c-2)}e^{-{2+c\over c}y_0}a^{-{2+c\over c}}.\label{largede}
\end{equation}
Since the universe is dominated by the dark energy for large $a$, we
have
\begin{equation}
\rho_\Lambda\simeq 3H^2={\rho_{\rm m}\over
1-\Omega_\Lambda}={\rho_{\rm m}^0a^{-3}\over 1-\Omega_\Lambda}.
\end{equation}
Thus, using Eq. (\ref{largede}) in the above relation, we derive
\begin{equation}
\rho_\Lambda=2^{-{2(2+c)\over 2-c}}(2+c)^{-{8\over
c(c-2)}}e^{{2+c\over c}y_0}\rho_{\rm m}^0a^{{2(1-c)\over c}}.
\end{equation}
It can be clearly seen from the above expression that the value of
$c$ plays a significant role in determining the final evolution of
the dark energy. When $c=1$, the holographic dark energy will become
a cosmological constant that is related to $\rho_{\rm m}^0$ through
the above relation without $a$. The choice of $c>1$ makes the
density of the dark energy continuously decrease, just like the case
of quintessence dark energy. When $c<1$, the density of dark energy
ceaselessly increases with the expansion of the universe. The
phantom behavior ($c<1$) results in the density of holographic dark
energy becoming enormously high when $a$ is very large. As the
universe goes into the high energy regime, quantum gravity begins to
operate, giving rise to the modified Friedmann equation,
(\ref{modiFeq}). However, it is very hard to justify the borderline
between the usual Friedmann equation and the modified Friedmann
equation. Namely, one cannot accurately say when the usual Friedmann
equation is replaced by the modified one. Hence, we have to set a
criterion by hand. One can assume that quantum gravity begins to
operate when $\rho_\Lambda=\eta\rho_{\rm c}$, where, say, $\eta\sim
{\cal O}(10^{-3})-{\cal O}(10^{-2})$. We thus derive, at that
moment, the scale factor of the universe,
\begin{equation}
a_\eta=\left[\eta 2^{2(2+c)\over 2-c}(2+c)^{8\over
c(c-2)}e^{-{2+c\over c}y_0}\left({\rho_{\rm c}\over\rho_{\rm
m}^0}\right) \right]^{c\over 2(1-c)}.\label{aeta}
\end{equation}

We assume that when $a>a_\eta$ the evolution of the universe is
governed by Eq. (\ref{modiFeq}). Here, any other forms of energy
have already been decayed away, and the dark energy is thus the
unique component of the universe; so we have
\begin{equation}
3H^2=\rho_\Lambda\left(1-{\rho_\Lambda\over\rho_{\rm
c}}\right).\label{modiFeq2}
\end{equation}
From this relation, we have
\begin{equation}
\Omega_\Lambda=1+{\rho_\Lambda\over \rho_{\rm c}-\rho_\Lambda}.
\end{equation}
This indicates that $\Omega_\Lambda>1$ when $a>a_\eta$, even though
the space of the universe is assumed to be flat. In concrete terms,
when $\rho_\Lambda\ll \rho_{\rm c}$, we have
$\Omega_\Lambda\rightarrow 1^+$; when
$\rho_\Lambda\rightarrow\rho_{\rm c}$, we have
$\Omega_\Lambda\rightarrow \infty$. And, as a contrast, see Eq.
(\ref{largede}), for large $a$ but $a<a_\eta$, we have
$\Omega_\Lambda\rightarrow 1^-$. Now the question naturally arises
of how to realize the $\Omega_\Lambda=1$ crossing. This question is
actually equivalent to the one of asking when the usual Friedmann
equation should be replaced by the modified one. The transition of
the two phases is ambiguous so that we have to set a connection when
$\rho_\Lambda=\eta\rho_{\rm c}$. Hence, the initial stage for Eq.
(\ref{modiFeq2}) is from $\Omega_\Lambda=1+\epsilon$, where
$\epsilon$ is a small positive number.

The modified Friedmann equation can be rewritten as
\begin{equation}
\tilde{h}^2={H^2\over \rho_{\rm c}}={\Omega_\Lambda-1\over
3\Omega_\Lambda^2},\label{tildeh}
\end{equation}
where the dimensionless parameter $\tilde{h}$ is positive for an
expanding universe and negative for a contracting universe. Here,
since an expanding universe is considered for illustration, we take
the positive value to $\tilde{h}$. Combining the definitions of
holographic dark energy and future event horizon, namely Eqs.
(\ref{de}) and (\ref{eh}), yields
\begin{equation}
\int_a^\infty {d\ln a'\over \tilde{h}a'}={c\over
\tilde{h}a\sqrt{\Omega_\Lambda}}.\label{rh2}
\end{equation}
Following Eq. (\ref{tildeh}), we have
\begin{equation}
{1\over \tilde{h}a}={\sqrt{3}\Omega_\Lambda\over
a\sqrt{\Omega_\Lambda-1}}.\label{fri2}
\end{equation}
Substituting (\ref{fri2}) into (\ref{rh2}) yields
\begin{equation}
\int_x^\infty dx' {\Omega_\Lambda\over
e^{x'}\sqrt{\Omega_\Lambda-1}}={c\sqrt{\Omega_\Lambda}\over
e^x\sqrt{\Omega_\Lambda-1}},\label{inteq}
\end{equation}
where $x=\ln a$. Then, taking the derivative with respect to $x$ in
both sides of this equation, one obtains a differential equation for
the fractional density of the holographic dark energy,
\begin{equation}
\Omega'_\Lambda=2\Omega_\Lambda(\Omega_\Lambda-1)\left({1\over
c}\sqrt{\Omega_\Lambda}-1\right),\label{deq2}
\end{equation}
where the prime denotes the derivative with respect to $\ln a$. This
differential equation governs the holographic evolution of the
universe for the high energy regime. Note that here $c<1$, and
$\Omega_\Lambda>1$; hence $\Omega'_\Lambda$ is always positive,
namely the fractional density of dark energy increases in time, the
correct behavior as we expect. This equation can be solved exactly,
and the solution is
\begin{equation}
\ln \sqrt{\Omega_\Lambda}+{c\over
2(1-c)}\ln(\sqrt{\Omega_\Lambda}-1)-{c\over 2(1+c)}\ln
(\sqrt{\Omega_\Lambda}+1)-{1\over
1-c^2}\ln(\sqrt{\Omega_\Lambda}-c)=\ln a+y_\eta,\label{sol2}
\end{equation}
where $y_\eta$ is an integration constant that can be determined by
an appropriate initial condition. Now let us deduce the equation of
state for the holographic dark energy in the high energy regime
($a>a_\eta$). Following the energy conservation equation of dark
energy, we have $w=-1-(1/3)(d\ln\rho_\Lambda/d\ln a)$. Writing
\begin{equation}
\rho_\Lambda=3\Omega_\Lambda \tilde{h}^2\rho_{\rm
c}={\Omega_\Lambda-1\over \Omega_\Lambda}\rho_{\rm c},
\end{equation}
one can easily obtain
\begin{equation}
w=-{1\over 3}\left(1+{2\over c}\sqrt{\Omega_\Lambda}\right).
\end{equation}
Interestingly, this relation is the same as in the low energy
regime; see Eq. (\ref{w}). Note that here $\Omega_\Lambda$ is
governed by Eq. (\ref{deq2}).

It has been pointed out that the borderline between the usual
Friedmann equation and the modified one is rather ambiguous. One has
to designate a contrived link condition; for example, one can assume
that the transition has occurred when $\rho_\Lambda=\eta\rho_{\rm
c}$, where $\eta\sim {\cal O}(10^{-3})-{\cal O}(10^{-2})$, say.
Therefore, the initial stage of the high energy regime is contrived
to be from $a=a_\eta$, where $\Omega_\Lambda=1+\epsilon$ with
$\epsilon=\eta/(1-\eta)$. At the moment of $a=a_\eta$ and
$\Omega_\Lambda\rightarrow 1^+$, the most important term on the
left-hand side of (\ref{sol2}) is the second term; we thus can
determine the integration constant:
\begin{equation}
y_\eta={c\over 2(1-c)}\ln {\eta\over 2(1-\eta)}-{c\over 2(1+c)}\ln
2-{1\over 1-c^2}\ln (1-c)-\ln a_\eta,\label{yeta}
\end{equation}
where $a_\eta$ is given by Eq. (\ref{aeta}). It is remarkable that
for a large $\Omega_\Lambda$, when the universe approaches the
turnaround point, the left-hand side of (\ref{sol2}) goes to zero.
This gives the scale factor corresponding to the turnaround,
\begin{equation}
a_{\rm max}=e^{-y_\eta},
\end{equation}
where $y_\eta$ is given by Eq. (\ref{yeta}). Namely, the maximum
scale factor (at the turnaround) of the universe in the cyclic
universe scenario with holographic dark energy is totally determined
by the constant $y_\eta$. It should be noted that the link condition
$\rho_\Lambda=\eta\rho_{\rm c}$ is rather important in this
scenario, even though it is somewhat artificial. The reason for
contriving such a link condition is that the borderline between the
classical and quantum gravities is not so clear. Next, let us give
several numerical examples. We take $c=0.8$, $\Omega_{\rm
m}^0=0.27$, and $h=0.72$ (here $h$ is the dimensionless Hubble
parameter of today), which gives rise to $y_0=-1.54$. The critical
density of the universe depends on the theory we use; if we use an
effective theory of loop quantum cosmology, we have $\rho_{\rm
c}\approx 0.82\rho_{\rm pl}=1.82\times 10^{76}~{\rm GeV}^4$; if we
use a braneworld scenario, we can treat the value of $\rho_{\rm c}$
from ${\rm TeV}^4$ to $10^{76}~{\rm GeV}^4$. Here we take the loop
quantum cosmology for illustration. The present density of dust
matter is $\rho_{\rm m}^0=1.13\times 10^{-47}~{\rm GeV}^4$. Then, we
can determine the values of $a_\eta$ and $a_{\rm max}$ if the value
of $\eta$ is given. We only show two examples, $\eta=10^{-3}$ and
$\eta=10^{-2}$. The choice of $\eta=10^{-3}$ gives
$a_\eta=2.86\times 10^{240}$ and $a_{\rm max}=1.53\times 10^{245}$;
the choice of $\eta=10^{-2}$ results in $a_\eta=2.86\times 10^{242}$
and $a_{\rm max}=1.50\times 10^{245}$.

In fact, this scenario has a fatal flaw, namely, a cyclic universe
has no a future event horizon in principle, since an observer can
eventually see the whole universe due to the cyclicity of the
universe, if he/she waits a sufficiently long time. Therefore, the
calculations in this paper break down in this regard. However, we
can rescue the model by reconsidering the IR cutoff of the universe.
In the original work of the holographic dark energy
\cite{Li:2004rb}, the choice of the future event horizon as an IR
cutoff is only a conjecture for ensuring the acceleration of the
universe. Now that the future event horizon cannot exist in a cyclic
universe, we might as well make a modification to the future event
horizon. We can define a ``finite-future event horizon'', by
replacing the infinity with a time $T$ in the upper limit of
integration in Eq. (\ref{eh}), where $T$ denotes the time of the
turnaround. Choosing the finite-future event horizon as the IR
cutoff of the universe undoubtedly makes the holographic dark energy
meaningful in the cyclic universe, at least in the expanding stage.
However, for the contracting stage of the cyclic universe, the IR
cutoff is ambiguous for us. We can steer clear of this difficulty by
considering the contracting stage as a time-reversal course of the
expanding stage. It should be admitted that this assumption is
rather strong. It should also be noted that the emphasis of this
paper is placed on the holographic evolution in the high energy
regime in the expanding branch and on how to link the low and high
energy regimes.

When the holographic phantom density reaches the critical density
$\rho_{\rm c}$, the universe starts to turn around and contract. In
the contraction phase, the physical rules are assumed to be totally
the same as in the expansion phase, i.e., the high energy regime is
governed by Eq. (\ref{modiFeq2}) and the low energy regime is
dictated by Eq. (\ref{Feq}). The only difference is that the Hubble
parameter has a negative value, but this does not affect the
evolutionary rules of the holographic dark energy; i.e., the
dynamical evolution of the holographic dark energy is still
controlled by the differential equations (\ref{deq}) and
(\ref{deq2}). As the universe contracts, at first the energy density
of the universe decreases, because the holographic phantom density
decreases in importance (then, the quintom behavior makes the
holographic dark energy become a quintessence-like component whose
density increases as the universe contracts, although not as
strongly as matter and radiation do), but then it again increases as
matter and radiation become dominant. Eventually, the energy density
reaches the high values at which the modifications to the Friedmann
equation become important. Once the energy density again hits the
same critical density $\rho_{\rm c}$, the universe stops
contracting, bounces, and once again expands. The bounce looks like
a ``big bang'' for us, and at this point the universe has its
smallest extent (smallest scale factor $a_{\rm min}$) and largest
energy density (the critical density $\rho_{\rm c}$). An
inflationary period may occur if the inflaton field can be excited
by some mechanism that can solve the flatness and horizon problems,
etc., and this can also generate the scale-invariant primordial
perturbations seeding the structure formation. For the inflationary
universe in a loop quantum cosmology, see, e.g., Refs.
\cite{loopinfl}. As the universe expands, its density deceases, and
it goes through the radiation dominated and matter dominated
periods, with the usual primordial nucleosynthesis, microwave
background, and large structure formation. Around a redshift $z\sim
{\cal O}(1)$, the universe begins to accelerate due to the existence
of dark energy (the holographic dark energy in this paper). The
holographic dark energy with $c<1$ can help realize the turnaround
discussed above. It is noteworthy that the cyclic universe discussed
in this paper is an ideal case, and there are still several severe
obstacles existing in the cyclic cosmology, such as the density
fluctuation growth in the contraction phase, black hole formation,
and entropy increase, which can obstruct the realization of a truly
cyclic cosmology. These problems are not addressed in this paper.

To summarize, in this paper we investigated the holographic dark
energy in a cyclic universe. We generalized the model of holographic
dark energy proposed in Ref. \cite{Li:2004rb} to the case of a
cyclic universe, and we studied the cosmological evolution of the
holographic dark energy in such a universe in detail. The
holographic dark energy with $c<1$ can realize a quintom behavior;
namely, it evolves from a quintessence-like component to a
phantom-like component. The phantom energy density will become very
large in the far future, which leads to a ``big rip'' singularity at
a finite time in which all bounded objects are finally disrupted.
However, when the energy scale becomes enormously large, quantum
gravity effects may bring significant modifications to the Friedmann
equation, leading to the possibility of the avoidance of
singularities. A modified Friedmann equation, (\ref{modiFeq}), along
with a phantom component and other matter components can realize a
cyclic universe scenario in which the cosmic evolution is
nonsingular. In such a cyclic scenario, the large densities cause
the universe to bounce when it is small, and to turn around when it
is large. We investigated the holographic phantom cosmological
evolution ($c<1$) in such a cyclic universe in detail. The dynamical
evolution of holographic dark energy in the low energy regime and in
the high energy regime are rather different; they are governed by
two differential equations, respectively. Linking the two regimes
together is a very important mission for this scenario. We proposed
a link condition connecting the regimes of low energy and high
energy, which gives rise to a complete picture of the holographic
evolution of the cyclic universe.

\section*{Acknowledgements}


We are grateful to Rong-Gen Cai, Hui Li and Zhi-Guang Xiao for
helpful discussions. This work was supported by the China
Postdoctoral Science Foundation, the K. C. Wong Education Foundation
(Hong Kong), the National Natural Science Foundation of China, and
the National Basic Research Program of China (2003CB716300).


\end{document}